\documentstyle[preprint,prd,aps]{revtex}
\begin{document}
\tightenlines

\title{Can thermal inflation solve the monopole problem ?}
\author{R. Jeannerot\\
{\em Centre for Theoretical Physics, University of Sussex,}\\
{\em Falmer, Brighton BN1 9QH, UK}}
\maketitle

\begin{abstract}
It is shown that thermal inflation arises naturally in rank greater than five 
unified theories when non-renormalisable terms are introduced. Thermal inflation 
is driven by two Higgs fields $\Phi_{B-L}$ and $\overline{\Phi}_{B-L}$ which 
also break $U(1)_{B-L}$ when acquiring vevs at the end of inflation. The 
inflationary period provides enough e-foldings to solve the monopole problem 
for $M_{B-L} \geq 10^{12}$ GeV. We point out that observations suggest that 
$M_{B-L} \simeq 10^{14}$ GeV.
\end{abstract}

\section{Introduction}
\label{sec-intro}

Supersymmetric grand unified theories (GUTs) of the strong, weak and 
electromagnetic interactions have received lots of interests since it has been 
shown that the three gauge coupling constants of the minimal supersymmetric 
standard model (MSSM), when 
interpolated to high energies, meet in a single point at $2 \times 10^{16}$ GeV 
\cite{meet}. GUT theories can also be 
viewed as low energy limits of a more fundamental theory. If this more 
fundamental theory is superstrings theory, GUTs could explain the discrepancy 
which exists between the string scale and the scale at which all fundamental 
interactions other than gravity merge. Therefore the idea of 
GUT is an idea which stands on its own, whether we believe or not in the 
existence of an ultimate theory of nature; and all problems which GUT 
theories have to deal with, such as the monopole problem, still have to be 
solved. If we do believe in the existence of an ultimate theory, which is 
probably a theory of quantum gravity , 
then we may want to include Planck suppressed operators (or operators suppressed 
by some very high energy scale) when building a GUT model. These operators then 
underline the fact 
that the GUT emerges from a more fundamental theory, but without any assumption 
upon 
the nature of this more fundamental theory. A problem which all GUTs have to 
face, whether they are based on 
semi-simple or non-semi-simple gauge groups, is the monopole problem. We 
investigate in this paper the possibly of the GUT itself solving its own 
monopole problem, by the so-called thermal inflation \cite{LythStew} which can 
emerge when Planck suppressed operators are introduced. We call 
these types of models self-consistent.

Thermal inflation was originally invented to solve the moduli problem which 
arises in superstrings theories \cite{LythStew}. However, it has been recently 
shown that this does not always work, in particular in the case of very light 
moduli\cite {Kim}. However, thermal inflation is a very interesting mechanism; 
we thus want to investigate the possibility of solving the monopole problem. We 
shall be looking at a particularly 
interesting set of GUT models 
based on rank greater than five theories such as SO(10), E(6), of GUT not based 
on semi-simple group such as the trinification $SU(3)_c \times SU(3)_L \times 
SU(3)_R$ or the Pati-Salam unified group $SU(4)_c \times SU(2)_L \times 
SU(2)_R$. We shall assume that there is an intermediate left-right symmetry, 
whose energy scale scale will be constrained from the conditions for successful 
thermal inflation to solve the monopole problem. We will be mainly interested in 
SO(10) GUT, the smallest GUT based on a 
semi-simple gauge group which unifies all fermions, including a right-handed 
neutrino. SO(10) has many desirable features, since it can explain the small 
neutrino masses which have recently been 
observed at SuperKamiokande \cite{Pati}, it can predict fermion textures 
\cite{fermions}, it also gives natural solutions to the doublet-triplet 
splitting 
problem \cite{babubarr,babumoha}. From a cosmological point of view, SO(10) 
predicts the existence of superheavy magnetic monopoles which have to be diluted 
if the theory is to be viable. 
It also predicts the existence of $B-L$ cosmic strings which can explain the 
baryon asymmetry which is observed in our universe \cite{paperlept}. 
It has also been shown that it is a-priori possible to get an SO(10) gauge 
symmetry at the GUT scale, 
with non-minimal Higgs structure, with arbitrary low left/right symmetry 
breaking scale \cite{Rizzo}.

In this paper we show that in GUT models based on rank greater than five 
gauge groups a period of the so-called thermal inflation \cite{LythStew} 
arises naturally if non-renormalisable term are introduced. If there is an 
intermediate left-right symmetry, the grand unified monopoles which are
previously formed can be diluted by the thermal inflation, such as to 
be consistent with the observational bound. At the end of inflation, 
$B-L$ fat strings are produced. 

In Sec.\ref{sec-mon} we review the monopole problem.

In Sec.\ref{sec-hybrid} we define what we call self-consistent models. In 
Sec.\ref{sec-LR}, we review the phenomenology of GUT models which break down to 
the Standard Model via an intermediate Left-Right gauge symmetry.  We show that 
such models can be made self-consistent.

In Sec.\ref{sec-thermal}, we show that thermal inflation arises naturally in 
rank greater than five GUTs if non renormalisable operators are included in the 
field sector which is used to lower the rank of the group by one unit, breaking 
$U(1)_{B-L}$. The inflationary period is driven by two inflaton fields. In 
Sec.\ref{sec-decay}, we use the vacuum expectation value (vev) of the inflaton 
fields, which transform as SU(5) singlets, to generate the $\mu$-term of the 
MSSM. In sec.\ref{sec-dilution}, we calculated the dilution factor which is 
obtained for various models. Cosmological constraints on the monopole abundance 
discussed in Sec.\ref{sec-mon} in turn lead to constraints on the $B-L$ breaking 
scale.

Finally, in Sec.\ref{sec-concl}, we conclude.

\section{The monopole problem}
\label{sec-mon}

In this section, we wish to briefly review the GUT monopole problem. Recall that 
monopoles are point like topological objects which form at phase transitions 
when the vacuum manifold contains non-contractible 2-spheres. Monopoles are very 
heavy objects, and because they are too many, they would dominate the energy 
density of the universe soon after formation. Therefore some mechanism has to be 
invoked to dilute them.

The monopole problem does not only arise in grand unified theories
based on semi-simple gauge group such as SU(5), SO(10) or E(6), but it 
also arises in partial unification theories such as those based on the
Pati-Salam group $SU(4)_c \times SU(2)_L \times SU(2)_R$ or the trinification
based on the $SU(3)_c \times SU(3)_L \times SU(3)_R$ gauge symmetry. In fact it 
arises in all extensions 
of the particle physics Standard Model as soon as the $U(1)_Y$ gauge symmetry
of the Standard model is embedded in a non abelian gauge symmetry which breaks 
at high energy. 

This can be understood as follows. Monopoles form according to the
Kibble mechanism at phase transitions 
associated with the spontaneous symmetry breaking of a group G down to a 
subgroup H of G if the second homotopy group of the vacuum manifold is non
trivial $\pi_2({G\over H}) \neq I$.
Now $\pi_1({G\over H}) \cong {\pi_1 (H_0) \over\pi_1 (G_0)} $ where $H_0$
($G_0$) is the component of H (G) connected to the identity. 
If we identify H with the standard model gauge group, $\pi_1 (H) =
\pi_1 (U(1)) = Z$ and hence topological monopoles form when G breaks down 
to H if $\pi_1 (G) = I$. This is always possible if we choose to work with the 
universal covering groups \cite{Tom}. To check whether the monopoles are 
topologically stable 
we must evaluate $\pi_2({G\over K})$ where $K= SU(3)_c \times U(1)_Q$. If 
$\pi_1 (G) = I$, $\pi_2({G\over K}) = \pi_1 (K) =Z$ and hence the
monopoles are topologically stable down to low energies.

Undependently of the initial monopole density, monopole-antimonopole 
pairs annihilate until the monopole-to-entropy ratio reaches its  final 
value \cite{Preskill79,ShelVil}:
\begin{equation}
{n_M\over s} \sim {1\over h^6 \beta \sqrt{g_{*s}}} {m \over M_{pl}}
\end{equation}
where $n_M$ is the monopole density and $s$ is the entropy. $h$ is the monopole 
magnetic charge which is given by  $h=-{4 
\pi \over e}$, where $e$ is the gauge coupling constant, $\beta \sim (1-5) g_*$, 
where $g_*$ is the effective number of degrees of freedom and $g_{*s}$ is the 
effective number of helicity states for particles with  mass $m+p < T$. Finally, 
$m$ 
is the monopole mass and $M_{pl}$ is the Planck mass. The monopole mass is 
bounded from below by \cite{Bogomol}:
\begin{equation}
m \geq {4 \pi \over e} T_M
\end{equation}
where $T_M$ is the temperature at which the monopoles form. For monopoles forming 
at the GUT scale $M_{GUT} \simeq 2\times 10^{16}$ GeV in a supersymmetric grand 
unified theory,  we find:
\begin{equation}
{n_M\over s} \sim 10^{-13}. \label{eq:mon}
\end{equation}
However, observations show that the monopole density today should be much 
smaller 
than that predicted by Eq.(\ref{eq:mon}). In fact, the strongest bound on the 
monopole density today comes neutron 
stars. Indeed, neutron stars 
could in principle trap monopoles \cite{ShelVil,KT}. Since we know that 
monopoles can catalyse proton decay, these would then increase the star 
luminosity. Limits 
on the luminosity of neutron stars imply a bound on the monopole flux which 
translates into a bound on the monopole density given by \cite{KT}:
\begin{equation}
{n_M \over s} \leq 10^{-31}. \label{eq:monobound}
\end{equation}
 We therefore need a dilution factor given by:
\begin{equation}
\Delta_{NS} \sim 10^{18}\label{eq:DNS}
\end{equation}
for GUT scale monopoles. 

The bound given 
in Eq.(\ref{eq:monobound}) can be improved by yet six orders of magnitude if one 
takes into account monopoles captured by the star when it was on the main 
sequence, before it became a neutron star \cite{ShelVil,KT}. We call is the {\em 
strong} neutron star bound:
\begin{equation}
{n_M \over s} \leq 10^{-37}. \label{eq:monobound2}
\end{equation} 
In such a case, 
the dilution factor which is required to solve the monopole problem is:
\begin{equation}
\Delta_{MSNS} \sim 10^{24} . \label{eq:DMSNS}
\end{equation}

If one hopes monopoles to be observed, the monopole density should not be much 
smaller than that ones given in Eq.(\ref{eq:monobound2}). If one wants to be 
conservative, it could be close to that given in Eq.(\ref{eq:monobound}).

\section{Solving the monopole problem}

\subsection{Self-consistent unified theories}
\label{sec-hybrid}

One would like a theory which is cosmologically consistent by itself. 
If a theory produces some cosmologically
catastrophic objects such as superheavy monopoles, it is desirable that the 
solution to 
this problem comes from the theory itself. There are two such compelling 
mechanisms for the monopole problem to be solved. Firstly, there is 
the Langacker-Pi mechanism \cite{LangPi} which assumes that electromagnetism 
is broken for a while, such that cosmic strings form connecting 
monopole-antimonopole pairs; the whole system of monopoles connected by 
strings rapidly decays. This scenario relies on a specific choice of the 
spontaneous symmetry breaking pattern.  Secondly, there is inflation which is 
a period of very rapid expansion of the early universe. Inflation makes the 
monopole density very low. Inflation must be driven by a scalar field with a 
flat potential: the inflaton. In supersymmetric extensions of unified theories,
inflation may just be a consequence of model building \cite{RJ3} (with an 
appropriate choice for the initial conditions). The inflaton field can be 
identified 
with a Higgs field used to implement the spontaneous symmetry breaking 
pattern of the unified gauge group down to the standard model gauge group or 
with a scalar field singlet under this group; such 
singlets are sometimes needed to make some of the Higgs fields to get a vev. 
If there is no extra symmetry nor any extra field than those needed to implement 
the full spontaneous symmetry breaking pattern from the considered GUT gauge 
group down to $SU(3)_c \times U(1)_Q$, some set of appropriate initial 
conditions may be needed, we may say that inflation arises from the GUT itself; 
and if this period of inflation makes the monopole density below the 
observational bound, that the theory is self-consistent.

For example, GUT models based on rank greater than five gauge groups $G$ 
naturally lead to a period of hybrid inflation \cite{RJ3} when the 
superpotential used to lower the rank of the group is given by 
\cite{Cop,ShaDva}:
\begin{equation}
W = \alpha S \overline{\Phi} \Phi - \mu^2 S \label{eq:hybrid}
\end{equation}
where $S$ is a scalar field singlet under the considered gauge group, which we 
identify with the inflaton,  $\Phi$ and
$\overline{\Phi}$  and Higgs superfields in complex conjugate 
representations which lower the rank of $G$ when acquiring vev, $\alpha$ and 
$\mu$ are two positive constant and
$\mu\over \sqrt{\alpha}$ sets the symmetry breaking scale. The global 
minimum of the potential is at $S=0$ and $\langle|\overline{\Phi}|\rangle = 
\langle 
|\Phi |\rangle=
{\mu\over \sqrt{\alpha}}$. Note that the above superpotential is consistent with 
the set of continuous R-symmetries. In this scenario, the $\overline{\Phi}$ 
and  $\Phi$ fields are kept at the origin during inflation, see 
Refs.\cite{ShaDva,RJ3},
and pick up a vev at the end of inflation. Therefore the symmetry breaking 
induced by the $\Phi$ and $\overline{\Phi}$ field vevs takes place at the end of 
inflation. Hence it should not produce
monopoles. Monopoles can be produced at a previous phase transition. 
Studying the formation of topological defects in all possible spontaneous 
symmetry breaking patterns from $G$ down to the Standard Model, with the 
assumption that the rank of the group is lowered with a superpotential
given in Eq.(\ref{eq:hybrid}), allows us to select the only models consistent 
with observation (see Ref.\cite{RJ1} for an SO(10) example).

Now consider a general hybrid inflationary scenario where the inflaton field is 
identified with a scalar field singlet under G which is used to make some GUT 
Higgs field $\Phi$ (not necessarily the one which lowers the rank of the group) 
to 
acquire a vev. Assume also that the phase transition induced by the vev of 
$\Phi$ leads to the formation of monopoles. By a suitable choice of initial 
conditions, the Higgs field $\Phi$ can be non zero
during inflation (the symmetry is already broken during inflation) and hence no 
monopole form at the end of inflation. Such a model based on SU(5) GUT
has been constructed \cite{Covi}, where $\Phi$ is a 24-dimensional Higgs 
superfield whose vev breaks SU(5) down to the standard model gauge group. This 
theory is self-consistent.

In general, several Higgs fields in various representations of the considered 
unified gauge group $G$ are needed to implement the full spontaneous symmetry 
breaking pattern from $G$ down to the Standard Model. One could use one (some) 
of these Higgs fields to play the role of the inflaton(s) (we can have more than 
one inflaton field). If the role of the inflaton is played by a (or some) Higgs 
field(s) which do(es) not lead to the formation of monopoles (or domain walls) 
when acquiring vev, it can have zero value during inflation and only acquire a 
vev at the end of inflation. In that case, topological defects (if any) 
associated with the spontaneous symmetry breaking induced by the vev
of the inflaton form at the end of inflation. The monopoles must then 
be formed by other Higgs field(s) which must acquire a vev before the time 
needed for the inflation to dilute the monopoles. Now, if the inflaton is (one 
of) the Higgs field(s) which lead(s) to the formation of monopoles, it must be 
non vanishing during the inflationary period; this is possible with a suitable 
choice of the initial condition. 

In the thermal inflationary scenario \cite{LythStew}, the scalar field which 
drives inflation is hold at the origin by finite temperature corrections to 
the effective potential during inflation, and only acquires a vev at the end of 
inflation. Therefore topological defects associated with the vev of this field 
form at the end of inflation. The grand unified monopoles must then be formed at 
a previous phase transition, if thermal inflation has to solve the monopole 
problem.

To summarize, many GUT models can be called self-consistent if inflation emerges 
naturally from the theory itself; they nevertheless  have to face some 
fine-tuning, as do all models of inflation, which correspond to a suitable 
choice of initial conditions. However, if the considered theory emerges from a 
quantum cosmological period, the choice of chaotic initial conditions may be the 
one which faces the least problems of naturalness.

\subsection{Unified models with intermediate left-right symmetry}
\label{sec-LR}

The most simple realistic examples of theories beyond the standard model which 
are self-consistent are unified theories based on rank greater than five gauge 
groups $G$ 
which break down to the Standard Model via an intermediate left-right gauge 
symmetry 
\cite{RJ3,RJcosmo}. The superpotential which reduces the rank of the group 
must be of the form given by Eq.(\ref{eq:hybrid}) and chaotic initial conditions 
have to be 
imposed. In 
this section, we wish to briefly review the phenomenology of such models and the 
steps for model building. In the next section, we will use the same spontaneous 
symmetry breaking pattern and the same set of Higgs fields to get a period of 
thermal inflation. It will be just a question of changing the form of the 
superpotential which breaks the $U(1)_{B-L}$ gauge symmetry.

Specifically, the spontaneous symmetry patterns that we consider are the 
following:
\begin{eqnarray}
G\times{\rm SUSY} \stackrel{M_{GUT}}\rightarrow ... &
 \stackrel{M_G}\rightarrow  &SU(3)_c \times SU(2)_L \times SU(2)_R \times 
U(1)_{B-L} \times{\rm 
SUSY}\nonumber\\
& \stackrel{M_R}\rightarrow & SU(3)_c \times SU(2)_L \times U(1)_R \times 
U(1)_{B-L} \times{\rm SUSY}\nonumber\\
& 
\stackrel{M_{B-L},\langle\Phi_{B-L}\rangle,\langle\overline{\Phi}_{B-L}\rangle}
\rightarrow& SU(3)_c \times SU(2)_L \times U(1)_Y (\times Z_2) \times{\rm 
SUSY}\nonumber\\
& \stackrel{M_R}\rightarrow & SU(3)_c \times U(1)_Q (\times Z_2)\label{eq:SSB}
\end{eqnarray}
where $G$ is a rank greater than five gauge group (not necessarily a 
semi-simple group), and
all cases with or without the intermediate gauge symmetry at 
$M_R$ or $M_{B-L}$ are relevant. In the latter case, we shall 
replace $M_R$ by $M_{B-L}$. $G$ can also break directly to the left-right gauge 
symmetry; this is the case where $M_{GUT} = M_G$. In Eq.(\ref{eq:SSB}), the 
discrete $Z_2$ symmetry which can be left 
unbroken at low energy if safe Higgs representations of G (who carry even $B-L$ 
charge) are used to break G down to the standard model gauge group 
\cite{Martin} plays the role of matter parity and can be identified with 
R-parity. In such models, dimension five operators of the MSSM which lead to 
rapid proton decay are forbidden and the LSP is stable. 

If we study the 
formation of topological defects in the symmetry breaking pattern 
(\ref{eq:SSB}), we find that topologically stable monopoles 
form at $M_G$ (or above, where appropriate), more monopoles form at $M_R$ and 
cosmic strings form at $M_{B-L}$. Therefore a period of inflation is need 
between $M_G$ ($M_R$) and $M_{B-L}$.

In what follows, we shall often identify $G$ with SO(10) which is the simplest 
rank five GUT based on a semi-simple gauge group.

 In supersymmetric theories, in order to break $B-L$ we need two Higgs fields in 
non trivial complex
conjugate representation of $G$, which we call $\Phi_{B-L}$ and
$\overline{\Phi}_{B-L}$. The flatness condition for the D-term leads to the 
useful relation 
$\langle|\Phi_{B-L}|\rangle=\langle|\overline{\Phi}_{B-L}|\rangle$. The 
component of $\Phi_{B-L}$ (and $\overline{\Phi}_{B-L}$) which gets a vev must be 
an SU(5) singlet. It may transform under $SU(2)_R$ as a doublet or as a triplet. 
In the case of $SO(10)$, this corresponds to $\Phi_{B-L}$ and
$\overline{\Phi}_{B-L}$ being identified with a $16$ and a $\overline{16}$ 
dimensional Higgs representations or a $126$ and a $\overline{126}$ dimensional 
representations respectively. If the component of $\Phi_{B-L}$ which acquires a 
vev transforms as 
a triplet under $SU(2)_R$, the 
discrete $Z_2$ symmetry is broken, and R-parity is not conserved. Some 
additional R-parity has to be imposed, or the yukawas couplings of the
dimension five operators leading to rapid proton decay must be extremely small. 
 On the other hand, if the component of $\Phi_{B-L}$ transforms as 
a triplet under $SU(2)_R$, the discrete $Z_2$ symmetry remains unbroken in the 
$SU(3)_c \times SU(2)_L 
\times U(1)_Y$ phase, R-parity is conserved and the LSP is stable. Also the 
right-handed neutrinos can acquire a superheavy Majorana mass via the following 
coupling:
\begin{equation}
\lambda_R \overline{\Phi}_{B-L} N N  \label{eq:NR}
\end{equation}
where $N$ is the right-handed neutrino field. In such a case, left-handed 
neutrinos acquire a mass via the see-saw mechanism \cite{seesaw}. Note that a 
Majorana mass for 
the right-handed neutrinos can also emerge in the case of an $SU(2)_R$ doublet, 
by introducing the following  non-renormalisable interaction:
\begin{equation}
\lambda_R \overline{\Phi}_{B-L}{\overline{\Phi}_{B-L}\over M_{pl}} N N 
.\label{eq:NRNR}
\end{equation}

In SO(10), all fermions
belonging to a single family, including a right-handed neutrino, are assigned 
to the 16 dimensional spinorial representation. Majorana masses for the 
right-handed neutrinos are thus possible via the following couplings:
\begin{equation}
\lambda_R^{ij} 16_i 16_j \overline{126}
\end{equation}
if a pair of ($126$ + $\overline{126}$) dimensional Higgs representations are 
used to break $U(1)_{B-L}$ and via:
\begin{equation}
\lambda_R^{ij} 16_i 16_j{ \overline{16}_H 16_H \over M_{pl}}
\end{equation}
if a pair of ($16$ + $\overline{16}$) dimensional Higgs representations are 
used. Note that it seems impossible to get a 126 dimensional Higgs 
representation  from string theory \cite{Dienes,Dienes2}. Therefore, we see that 
both 
the $16$ + $\overline{16}$ and the $126$ + $\overline{126}$ dimensional 
representations have their own advantages and their own disadvantages.
For the purpose of thermal inflation both a pair of $16$ + $\overline{16}$ or a 
pair of
$126$ + $\overline{126}$ dimensional representations can be used.

To break G down to the standard model gauge group, we need extra Higgs on top of 
the  $\Phi_{B-L}$ and $\overline{\Phi}_{B-L}$ fields. In the case of SO(10) for 
example, we need some Higgs in the adjoint representation (the 45 dimensional 
representation) and some 
Higgs in the 54 dimensional representation. The number of extra Higgs  
depends on the number and/or on the nature of intermediate symmetry that we 
require in Eq.(\ref{eq:SSB}), ie. whether we want both $3_c \, 2_L \, 2_R \, 
1_{B-L}$ and  $3_c \, 2_L \, 1_R \, 1_{B-L}$, only $3_c \, 2_L \, 2_R \, 
1_{B-L}$ or only $3_c \, 2_L \, 1_R \, 1_{B-L}$. We do not consider them here, 
since they are irrelevant for our analysis. But we suppose that
if they couple to the $16$ + $\overline{16}$ ($126$ + $\overline{126}$) sector, 
the couplings are such that as not to destabilise the required vevs.

The simplest superpotential involving the $\Phi_{B-L}$ and 
$\overline{\Phi}_{B-L}$ fields which lowers the rank of the group by one unit, 
breaking $U(1)_{B-L}$, is given by Eq.(\ref{eq:hybrid}) if we identify the 
$\Phi$ and 
$\overline{\Phi}$ fields with $\Phi_{B-L}$ and $\overline{\Phi}_{B-L}$ 
respectively. 
The ratio $\mu \over \sqrt{\alpha}$ then sets the $B-L$ breaking scale, 
$M_{B-L}$. It 
is well known, as mentioned in the previous section, that this superpotential, 
with appropriate choice of initial conditions, leads to inflation 
\cite{Cop,ShaDva,RJ3}. Hence monopoles which are previously formed (see above) 
are inflated away. At the end of inflation, both $\Phi_{B-L}$ and 
$\overline{\Phi}_{B-L}$ acquire a vev, and $B-L$ cosmic strings form. They are 
called $B-L$ cosmic strings because the Higgs fields which form the strings is 
the same Higgs fields which are used to break 
$U(1)_{B-L}$. These cosmic strings can generate the baryon asymmetry which is 
observed in our universe \cite{paperlept}. The scenario takes place at end 
inflation.

The shape of the power spectrum of the cosmic microwave background (CMB) could 
in principle distinguish between different models of inflation. In 
such hybrid inflationary scenarios both strings and inflation contribute to the 
CMB anisotropies, and the string contribution could be as high as 75\% 
\cite{RJ3}. In the hybrid scenario, 
$M_{B-L}$ is constrained by COBE data to be $M_{B-L} \simeq 4.7 \times 10^{15}$ 
GeV \cite{ShaDva,RJ2,RJ3}. Although no full 
computation of the CMB power spectrum for a mixed scenario with inflation and 
cosmic strings has been done yet, because this first 
requires a better understanding of a cosmic string network evolution in an 
expanding universe \cite{Joao}, one can hope that by the time satellites like 
MAP or Planck will be launched, a full understanding of the strings evolution 
will be reached, and the computation of the CMB power spectrum in theories with 
both strings and 
inflation will be done. Since no scenario up to date does not seem to fit 
perfectly the CMB data, a mixed scenario with inflation and cosmic strings may 
well be the perfect fit! However, in these hybrid models, the strings forming at 
the end of inflation are very heavy. If we believe the recent string simulation 
of Vincent et 
al.\cite{Vincent} which have shown that the main energy loss mechanism for 
cosmic strings is via particle emission rather than gravitational radiation, 
the non observation of HECR above $10^{20}$ eV can rule out 
the existence of such heavy strings \cite{Vincent}. 

Now, if instead the $B-L$ strings where fat strings, ie arising from a 
superpotential which is flat in the $\Phi_{B-L}$ and in the 
$\overline{\Phi}_{B-L}$ directions, with a TeV mass Higgs and a very heavy gauge 
boson, they could explain the extra galactic diffuse 
$\gamma$-ray background above $\sim 10$ GeV, together with the highest energy 
cosmic ray flux above $\sim 10^{11}$ GeV \cite{HECR}. They would have to form  
at $T \sim 
10^{14}$ GeV.  We shall keep in mind the value of 
this very interesting energy scale of $10^{14}$ GeV for $B-L$ string formation.

Finally, in order to break $U(1)_{B-L}$, instead of introducing a scalar field 
singlet under $G$ to help the 
$\Phi_{B-L}$ and $\overline{\Phi}_{B-L}$ fields to get a vev, we could use 
a superpotential which involves non-renormalisable terms. This is what we shall 
do in the next section, and see that it may lead to a period of thermal 
inflation which can make the monopole density in agreement with observation. At 
the end of this thermal inflationary period, fat $B-L$ cosmic strings form.

\section{Introducing non-renormalisable terms}

Unified theories of the strong, weak and electromagnetic interactions are 
probably the remnants of a more fundamental theory of nature only valid above 
some 
very high energy threshold, possibly the Planck scale. This ultimate theory of 
nature 
could manifest itself in the low energy world through Planck suppressed 
interactions. Note that we do not assume that this ultimate theory is a theory 
of superstrings, but we do assume that this theory shares similar properties 
with superstring theory, and in particular that it may lead  to 
interactions which are suppressed by some very high scale which we take to be 
the Planck scale for simplicity. We therefore now turn to the second possibility 
for building a 
superpotential in 
the $B-L$ sector, see Sec.\ref{sec-LR}, which involves non-renormalisable terms. 
With appropriate 
choice of initial conditions and a set of continuous R-symmetries, we show that 
a period of thermal inflation emerges which solves the monopole problem. 

\subsection{Thermal inflation and the breaking of $U(1)_{B-L}$}
\label{sec-thermal}

Here again we consider unified models based on a rank greater than five gauge 
group
with a spontaneous symmetry breaking pattern of the form given by 
Eq.(\ref{eq:SSB}). The simplest superpotential involving non-renormalisable 
terms 
and which leads to the spontaneous symmetry breaking of $U(1)_{B-L}$,  involving 
the $\Phi_{B-L}$ and $\overline{\Phi}_{B-L}$ fields described in 
Sec.\ref{sec-LR}, is given by:
\begin{equation}
W = \lambda {(\overline{\Phi}_{B-L} \Phi_{B-L})^2 \over M_{pl}}  
\label{eq:super}
\end{equation}
where $M_{pl}$ is the Planck mass and $\lambda$ is a positive coupling constant. 
Eq.(\ref{eq:super}) must involve all dimension 4 invariant terms which 
can be made using  the $\Phi_{B-L}$ and 
$\overline{\Phi}_{B-L}$ fields. For example, in the case of SO(10),  
when the $\Phi_{B-L}$ and $\overline{\Phi}_{B-L}$ fields are a pair of $126$ + 
$\overline{126}$ dimensional representations, the superpotential is given 
by \cite{babubarr}:
\begin{equation}
W = \lambda {(\overline{\Phi}_{B-L} \Phi_{B-L})^2 \over M_{pl}} + \alpha 
{(\overline{\Phi}_{B-L} \gamma^{ab} \Phi_{B-L})^2 \over M_{pl}} 
\end{equation}
where $\gamma^{ab}$ are generalised gamma matrices.
As mentioned before, we need more Higgs to complete the spontaneous symmetry 
breaking of SO(10) down to the standard model gauge group, such as 45's 
dimensional representations or 54's dimensional ones. We do not consider them 
here, since they are irrelevant for our analysis. 

The effective scalar potential for the flaton fields  $\Phi_{B-L}$ and
$\overline{\Phi}_{B-L}$ including the soft supersymmetry breaking terms is given 
by:
\begin{eqnarray}
V &=& V_0 + 4 \lambda^4 {|\overline{\Phi}_{B-L}^4 \Phi_{B-L}^2| \over M_{pl}} + 
4 \lambda^4 {|\overline{\Phi}_{B-L}^2 \Phi_{B-L}^4| \over M_{pl}} + {g^2 q^2  
\over 2} (|\Phi_{B-L}|^2 - |\overline{\Phi}_{B-L})^2|^2\nonumber\\
&&- m_{\Phi_{B-L}}^2 \Phi_{B-L}^2 - m_{\overline{\Phi}_{B-L}}^2 
{\overline{\Phi}_{B-L}}^2  + \lambda A {(\overline{\Phi}_{B-L} \Phi_{B-L})^2 
\over M_{pl}} + c.c. \label{eq:pot}
\end{eqnarray}
where we have used the same notation for the Higgs superfields and their bosonic 
components. $m_{\Phi_{B-L}}$, $m_{\overline{\Phi}_{B-L}}$ and $A$ are soft 
supersymmetry 
breaking parameters. The charge $q =1$ for the SU(5) singlet component of 
$\Phi_{B-L}$ 
transforming as an 
$SU(2)_R$ doublet and $q=2$ if it transforms as an $SU(2)_R$ triplet (see 
Sec.\ref{sec-LR}). $V_0$ will 
be determined by the requirement that the cosmological constant today vanishes.

We would like to point out that the $\Phi_{B-L}$ and $\overline{\Phi}_{B-L}$ 
fields do not 
necessarily have the same soft 
mass terms at low energy; $m_{\Phi_{B-L}} \neq m_{\overline{\Phi}_{B-L}}$ in 
general. This is because the $\Phi_{B-L}$ and $\overline{\Phi}_{B-L}$ fields do 
not couple 
to the same particles of the theory, and $m_{\Phi_{B-L}}$ and 
$m_{\overline{\Phi}_{B-L}}$ depend on all these 
couplings \footnote{One should in fact compute the renormalisation group 
equations to confirm both the sign and the magnitude of $m_{\Phi_{B-L}}$ and 
$m_{\overline{\Phi}_{B-L}}$; see for exemple \cite{Murayama92}, where it is 
shown that couplings of the Higgs particle to fermions drives its soft mass 
negative.}. The $\overline{\Phi}_{B-L}$ field couples 
to the right handed neutrinos, whereas the $\Phi_{B-L}$ field does not (see 
Sec.\ref{sec-LR}).  However, both  the $\Phi_{B-L}$ and $\overline{\Phi}_{B-L}$ 
fields also 
give a 
contribution to the mass of the $B-L$ gaugino\footnote{Our notation might be 
confusing, but we hope that it was obvious to the reader that when we say $B-L$ 
gaugino we really mean a gaugino which is a linear combination of the $U(1)_R$ 
and the $U(1)_{B-L}$ gauginos, in the same way that when we say $\Phi_{B-L}$ 
breaks $B-L$ we mean that it breaks a linear combination of $U(1)_R$ and 
$U(1)_{B-L}$, since the remaining gauge symmetry much be $U(1)_Y$, and that in 
theories such as SO(10) the hypercharge is given by $ {Y\over 2} = I_{3R} + 
{B-L\over 2}$.}. Furthermore, if we want both the $ \Phi_{B-L}$ and 
$\overline{\Phi}_{B-L}$ fields to acquire a vev, both $m_{\Phi_{B-L}}$ and 
$m_{\overline{\Phi}_{B-L}}$ must be negative with $m_{\Phi_{B-L}} = 
m_{\overline{\Phi}_{B-L}}$ if spontaneous symmetry breakdown in the D-flat 
direction is required. Thus the simplifying assumption $m_{\Phi_{B-L}} = 
m_{\overline{\Phi}_{B-L}}$ can be justified. 

We now turn to the finite temperature effective potential. Both the $\Phi_{B-L}$ 
 and $\overline{\Phi}_{B-L}$ fields couples to the
$\tilde{\Phi}_{B-L}$, $A_{B-L}$, $\tilde{A}_{B-L}$  fields and 
$\overline{\Phi}_{B-L}$ 
also the $N$ and $\tilde{N}$ fields. We have used a tilde to denote 
supersymmetric particles. Therefore the finite temperature effective potential 
can be 
calculated \cite{Tiago}. We have:
\begin{eqnarray}
V &=& V_0 + 4 \lambda^4 {|\overline{\Phi}_{B-L}^4 \Phi_{B-L}^2| \over M_{pl}} + 
4 \lambda^4 {|\overline{\Phi}_{B-L}^2 \Phi_{B-L}^4 |\over M_{pl}} + {g^2 q^2  
\over 2} (|\Phi_{B-L}|^2 - |\overline{\Phi}_{B-L})^2|^2\nonumber\\
&&+ (\alpha_{\Phi} T^2 - m_{\Phi_{B-L}}^2) \Phi_{B-L}^2 
(\alpha_{\overline{\Phi}} T^2- m_{\overline{\Phi}_{B-L}}^2) 
{\overline{\Phi}_{B-L}}^2  + \lambda A {(\overline{\Phi}_{B-L} \Phi_{B-L})^2 
\over M_{pl}} + c.c. \label{eq:finite}
\end{eqnarray}
where $\alpha_{\Phi} \sim \alpha_{\overline{\Phi}} \sim O(1)$.

 From the D-flat condition, we must have $\langle |\overline{\Phi}_{B-L}|\rangle 
= \langle |\Phi_{B-L}| \rangle$ which we call $\eta$. Now minimizing the finite 
temperature 
effective potential given by Eq.(\ref{eq:finite}) along the D-flat direction, we 
find 
that there are two possible minima:
\begin{enumerate}
\item For $T^2 \gg m_{\Phi_{B-L}} (= m_{\overline{\Phi}_{B-L}})$, the minima is 
at $\eta = 0$ and the symmetry is restored. 

\item For $T^2 < m_{\Phi_{B-L}}$, 
\begin{equation}
\eta = \left ( (\sqrt{13} + 1) {M_{pl} m \over 12 \lambda} \right )^{1\over 
2}\label{eq:vev}
\end{equation}
where $m \simeq 10^2 - 10^3$ GeV is the soft supersymmetry breaking scale 
in gravity mediated supersymmetry breaking scenarios and we have $m \sim m_\Phi 
\sim A$. In that case, the $U(1)_{B-L}$ gauge symmetry is spontaneously broken. 
\end{enumerate}
Thus at very high temperatures, the $\Phi_{B-L}$ and  $\overline{\Phi}_{B-L}$  
fields are kept at the origin by finite temperature corrections to the effective 
potential. The energy density of the universe is dominated by the 
vacuum energy density $V_0$, see Eq.(\ref{eq:pot}), and a period of inflation 
takes place. When the temperature $T$ falls below $\sim m_{\Phi_{B-L}}$, the 
inflationary period stops. $V_0$ can be determined by requiring that the 
cosmological constant today vanishes. We find: 
\begin{equation}
V_0^{1\over 4} \sim 2^{1\over 4}  m^{1\over 2} \eta^{1\over 2}. \label{eq:V0}
\end{equation}

We have thus shown that in these models a period of thermal inflation can take 
place \cite{LythStew}. This is driven by two inflaton fields, the $\Phi_{B-L}$ 
and  $\overline{\Phi}_{B-L}$  fields. The cosmological scenario is as follows. 
At very high 
temperatures, the 
$\Phi_{B-L}$ and $\overline{\Phi}_{B-L}$ fields are kept at the origin by the 
finite 
temperature corrections to the effective potential. The spontaneous symmetry 
breaking of $G$ down to $SU(3)_c \times SU(2)_L 
\times U(1)_R \times U(1)_{B-L}$ (or $SU(3)_c \times SU(2)_L 
\times SU(2)_R \times U(1)_{B-L}$, see Sec.\ref{sec-LR}) takes place and 
topologically stable monopoles form. 
The intermediate gauge symmetry remains unbroken as long as the finite 
temperature corrections are strong enough to keep the $\Phi_{B-L}$ and 
$\overline{\Phi}_{B-L}$ fields at the origin.  When the temperature falls below 
$\sim 
m_{\Phi_{B-L}}$, the 
SU(5) singlet components of $\Phi_{B-L}$ and $\overline{\Phi}_{B-L}$ acquire a 
vev, which is given by Eq.(\ref{eq:vev}); the intermediate gauge symmetry 
spontaneously breaks 
down to the Standard Model, and $B-L$ cosmic strings form. They are fat strings 
\cite{Tiago}. These $B-L$ cosmic strings which form at the end of inflation can 
generate the baryon asymmetry which is observed in our universe 
\cite{paperlept}. Alternatively, Affleck-Dine baryogenesis may take place 
\cite{SKY}.

\subsection{The decay of the inflaton fields}
\label{sec-decay}

We now need to find some decay channel for the inflaton field $\Phi_{B-L}$. 
Recall that the 
$\overline{\Phi}_{B-L}$ field gives a Majorana mass to the right-handed 
neutrinos and 
hence can decay into right-handed neutrinos, see Sec.\ref{sec-LR}. But the decay 
rate of 
$\overline{\Phi}_{B-L}$ into right-handed neutrinos is very small \cite{SKY}, 
and it also leads to a reheat temperature far below the required temperature 
which is 
needed not to overproduce LSP's \cite{SKY}. We thus follow the idea of 
Ref.\cite{SKY} and use the vevs of the $\Phi_{B-L}$ and $\overline{\Phi}_{B-L}$ 
fields which are SU(5) singlets to generate the $\mu$ term of the MSSM. 

We thus introduce the following 
superpotential:
\begin{equation}
W_\mu = \beta {(\overline{\Phi}_{B-L} \Phi_{B-L})^n \over M_{pl}^{2 n -1}}  H_u 
H_d \label{eq:mu}
\end{equation}
where $H_u$ and $H_d$ are the two Higgs doublets of the MSSM, which respectively 
give mass to up-type quarks and down-type quarks. $\beta$ is a coupling 
constant and $n$ is an integer. Recall that the superpotential which describes 
the MSSM is given by \cite{Nilles}:
\begin{equation}
W_{MSSM} = h_t Q U H_u + h_b Q U H_d + h_\tau L E H_d + \mu H_u H_d .
\label{eq:MSSM}
\end{equation}
We thus have:
\begin{equation}
\mu = \beta {(\langle\overline{\Phi}_{B-L}\rangle \langle 
\Phi_{B-L}\rangle)^n \over M_{pl}^{2n-1}}.\label{eq:muterm}
\end{equation}
The value of $n$ will be determined once the $B-L$ breaking scale $M_{B-L} 
\equiv \eta = \langle 
|\overline{\Phi}_{B-L} |\rangle$ for successfull inflation has been calculated.

In the case of SO(10), the $\mu$-term can arise from the following coupling:
\begin{equation}
W_\mu = \beta {(\overline{16}_H 16_H)^n \over M_{pl}^{2n-1}}  
10_H 10'_H
\end{equation}
provided that the doublet-triplet splitting has been solved via the mechanism of 
Ref.\cite{babumoha}. This requires the introduction of two Higgs multiplets in 
the $10$-dimensional representation of SO(10) and a Higgs in the adjoint 
representation with a vev in the $B-L$ direction, and mixing with the spinorial 
sector. The superpotential which leads to doublet-triplet splitting is given by 
\cite{babumoha}:
\begin{equation}
W = 10_H 45_H 10'_H + \lambda_1 16_H 16_H 10_H
+ \lambda_2 \overline{16}_H \overline{16}_H 
10'_H .
\end{equation} 
In that case, the two Higgs doublets of the MSSM belong to two different Higgs 
multiplets $10$ and $10'_H$.

If the doublet-triplet splitting has been solved using the 
Dimopoulos-Wilczek mechanism \cite{DW}, the $\mu$-term can then arise via the 
coupling:
\begin{equation}
W_\mu = \beta {(\overline{126}_H 126_H)^n \over M_{pl}^{2n-1}}  
H^2_{10}
\end{equation}
The Dimopoulos-Wilczek mechanism in SO(10) requires the introduction of 
a Higgs in the 45 dimensional representation with a vev in the $B-L$ 
direction and two Higgs in the 10 dimensional representation 
\cite{DW,babubarr}. The superpotential which then leads to doublet-triplet 
splitting is given by \cite{babubarr}:
\begin{equation}
W = 10_H 45_H 10'_H + M^2 10'_H
\end{equation} 
where $M$ is a superheavy mass scale. In that case, the two Higgs doublets of 
the 
standard model belong to the same Higgs multiplet $10_H$.

We are now ready to calculate the decay rates of the $\Phi_{B-L}$ and 
$\overline{\Phi}_{B-L}$ fields into standard model particles. By expanding the 
quantum 
field operators as:
\begin{eqnarray}
\hat\Phi_{B-L} &=& \Phi_{B-L} + \delta \Phi_{B-L} \\
\hat{\overline{\Phi}}_{B-L} &=& \overline{\Phi}_{B-L} + \delta 
\overline{\Phi}_{B-L}
\end{eqnarray} 
where $\Phi_{B-L} = \langle \Phi_{B-L} \rangle$ and $\overline{\Phi}_{B-L} = 
\langle \overline{\Phi}_{B-L} \rangle$, we can then deduce from 
Eqs.(\ref{eq:mu}) and 
(\ref{eq:MSSM}) the Lagrangian for the quantum fields. It is given by:
\begin{eqnarray}
- L_{decay} &\simeq& 2 \mu^2 (H_d^2 + H_u^2) \left( {\delta \Phi_{B-L} \over 
\eta} + {\delta \overline{\Phi}_{B-L} \over \eta}\right ) + 2\mu  h_t Q U H_u 
\left( {\delta \Phi_{B-L} \over \eta} + {\delta \overline{\Phi}_{B-L} \over 
\eta} \right ) \nonumber\\
&&+ 2 \mu \lambda_b Q U H_u \left( {\delta \Phi_{B-L} \over \eta} + {\delta 
\overline{\Phi}_{B-L} \over \eta} \right )+ 2 \mu \lambda_\tau L E H_d \left( 
{\delta \Phi_{B-L} \over \eta} + {\delta \overline{\Phi}_{B-L} \over \eta} 
\right )\nonumber\\
&& + 2 A \mu H_u H_d  \left( {\delta \Phi_{B-L} \over \eta} + {\delta 
\overline{\Phi}_{B-L} \over  \eta} \right ) - 4 \mu \tilde{H}_u \tilde{H}_d  
\left( {\delta \Phi_{B-L} \over \eta} + {\delta \overline{\Phi}_{B-L} \over  
\eta} \right ) \label{eq:Ldecay}
\end{eqnarray}
where we have written only leading order terms in $\Phi_{B-L}$ and 
$\overline{\Phi}_{B-L}$. From Eq.(\ref{eq:Ldecay}) we can deduce the decay 
rates for both the $\Phi_{B-L}$ and $\overline{\Phi}_{B-L}$ fields. They are 
estimated to be \cite{Linde}:
\begin{equation}
\Gamma_{\Phi_{B-L}} \sim \Gamma_{\overline{\Phi}_{B-L}} \sim {m^3 \over \pi 
\eta^2}\label{eq:rate}
\end{equation}
where we have assumed that $\mu \sim A \sim m_{\Phi_{B-L}} \sim 
m_{\overline{\Phi}_{B-L}} 
\sim m$. 

\subsection{The dilution factor}
\label{sec-dilution}

At the end of inflation, the two inflaton fields $\Phi_{B-L}$ and 
$\overline{\Phi}_{B-L}$ oscillate and rapidly decay into standard model 
particles, see Eq.(\ref{eq:Ldecay}), at the same decay rate which is given by 
Eq.(\ref{eq:rate}). 
This process reheats the Universe very quickly \cite{KT,Linde}.

By assuming that there is no parametric resonance, the reheating temperature at 
the end of inflation $T_R$ is related to the total decay width $\Gamma_{tot}$ by 
\cite{Linde}:
\begin{equation}
{\pi g_*(T_R) \over 30} T_R^4 \sim {\Gamma_{tot}^2 M_{pl}^2 \over 24} 
\label{eq:TR}
\end{equation}
where 
\begin{equation}
\Gamma_{tot} = \Gamma_{\Phi_{B-L}} + \Gamma_{\overline{\Phi}_{B-L}} \sim 2 
\Gamma_{\Phi_{B-L}} \label{eq:Gtot}
\end{equation}
and $g_*$ counts the number of massless degrees of freedom at $T_R$.

Now the dilution factor provided by the period of thermal inflation is given by 
\cite{LythStew}:
\begin{equation}
\Delta \sim {90 V_0 \over 3 \pi^2 g_*(T_c) T_c^3 T_R} \label{eq:Delta}
\end{equation}
where $T_c \sim m$ is the critical temperature at which the phase transition 
associated by the spontaneous symmetry breaking of $U(1)_{B-L}$ induced by the 
vevs of $\Phi_{B-L}$ and $\overline{\Phi}_{B-L}$ takes place.

Combining Eqs.(\ref{eq:Delta}),(\ref{eq:TR}),(\ref{eq:Gtot}), (\ref{eq:rate}) 
and  (\ref{eq:V0}), we can determine the $B-L$ breaking scale $M_{B-L} \equiv 
\eta$. We find:
\begin{equation}
M_{B-L} \sim \left ({\pi^{3\over2} \over 60}   g_*(T_c) \left ( {5\over \pi 
g_*(T_R)}\right )^{1\over 4} \Delta m^{5\over 2} M_{pl}^{1\over 2} \right 
)^{1\over 3} .
\end{equation}
In Table 1, we give the numerical values obtained for 
for the scale $M_{B-L}$, the vacuum 
energy $V_0^{1\over 4}$  during thermal inflation and
the couplings $\lambda$ and $\beta$,
for monopole densities satisfying the 
neutron stars limit 
given in Eq. (\ref{eq:monobound}) and the stronger limit given in 
Eq.(\ref{eq:monobound2}). $V_0^{1\over 4}$  is given in Eq.(\ref{eq:V0}), 
$\lambda$ appears 
in the superpotential given in Eq.(\ref{eq:super}) and $\beta$  appears in the 
$\mu$-term , see 
Eq.(\ref{eq:muterm}).
 We give results for two values of the soft supersymmetry 
breaking scale $m 
= 10^2$ GeV and $m=10^3$ GeV. We use the values $g_*(T_c) \sim 10^2$, $g_*(T_R) 
\sim 10$, 
$M_{pl} = 2.4 \times 10^{18}$ GeV.

\vspace{1cm}

\begin{tabular}{||p{1.1cm}@{}||p{3.2cm}@{}|p{3.2cm}@{}||p{3.2cm}|p{3.2cm}@{}||}
\cline{2-5} 
\cline{2-5}
\multicolumn{1}{c||}{ }&
\multicolumn{2}{|c||}{$\Delta_{NS} = 10^{18}$} & 
\multicolumn{2}{c||}{$\Delta_{MSNS} = 10^{24}$}\\ \hline\hline
$m$ & $10^3$ GeV & $10^2$ GeV &$10^3$ GeV &$10^2$ GeV \\\hline
$M_{B-L}$ & $6.6 \times 10^{11}$ GeV & $9.7 \times 10^{10}$ GeV & $6.6 \times 
10^{13}$ GeV& $9.7 \times 10^{12}$ GeV\\\hline
$V_0^{1\over 4}$ & $3.1 \times 10^7$ GeV & $3.7 \times 10^6$ GeV& $3.1 \times 
10^8$ GeV & $3.7 \times 10^7$ GeV\\\hline
$\beta$ & $5.5 \times 10^{-3}$ & $2.6 \times 10^{-2}$ & $5.5 \times 10^{-7}$& $ 
2.6 \times 10^{-6}$  \\\hline
$\lambda$ & $2.1 \times 10^{-3}$ & $9.8 \times 10^{-3}$ & $2.1 \times 10^{-7}$ & 
$9.8 \times 10^{-7}$ \\\hline
$m_{\nu L}^\tau$ & $15.15$ eV & $103.09$ eV& $0.15$ eV & $1.03$ eV \\ \hline 
\hline
\end{tabular}
\vspace{.5cm}

{\underline{Table 1:}} $\Delta_{NS}$ and 
$\Delta_{MSNS}$ are the dilution factors which are required from the 
neutron star bound and the {\em strong} neutron star bound respectively. $m$ is 
the soft supersymmetry breaking scale, $M_{B-L}$ is the $U(1)_{B-L}$ breaking 
scale, 
$V_0$ is the vacuum energy which dominates the energy density of the universe 
during thermal inflation, $\lambda$ is a Yukawa coupling which appears in 
Eq.(\ref{eq:super}), $\beta$ is a Yukawa coupling which is given by the 
$\mu$-term and  $m_{\nu L}^\tau$ is the tau neutrino mass.

\vspace{1cm}

All values for $\beta$ given in Table 1 correspond to $n=1$ in 
Eq.(\ref{eq:muterm}). In Table 1, we have also given the $\tau$ neutrino mass, 
assuming that the right-handed neutrino acquires its mass via the renormalisable 
coupling given by Eq.(\ref{eq:NR}). We thus have $m_\nu^\tau \simeq {100^2 \, 
{\rm GeV}^2 \over M_{B-L}}$. If neutrino masses arise from the 
non-renormalisable 
coupling given in Eq.(\ref{eq:NRNR}), we get tau neutrino masses far above the 
limits imposed by SuperKamiokande data \cite{Pati}. By assuming that 
SuperKamiokande observation represent $\nu_\mu \rightarrow 
\nu_{\tau}$-oscillations, 
the observed mass difference $\delta m^2 \simeq 10^{-2} - 10^{-3} \, {\rm 
eV}^2$, implies a tau neutrino mass  $m_{\nu L}^\tau \simeq {1\over 10} - 
{1\over 30}$ 
eV \cite{Pati}. This corresponds to $M_{B-L} \simeq 1 \times 10^{14} - 3 \times 
10^{14}$ which then gives a dilution factor $\Delta \simeq (0.3 - 9.4) \times 
10^{25}$ which is bigger than the strong observational limit given by neutron 
stars. Thus in such a scenario, the monopole problem is solved. Note that these 
values correspond to $\beta \simeq (2.4 -0.2) \times 10^{(-7)}$ and $\lambda 
\simeq (9.2 - 1.0) \times 10^{(-8)}$ (for $m = 10^3$ GeV). It is particularly 
interesting that this energy scale $\sim 10^{14}$ GeV required by neutrino 
masses observations, is the same energy scale 
which is needed for the fat $B-L$ cosmic strings to explain the extra galactic 
diffuse 
$\gamma$-ray background above $\sim 10$ GeV, together with the highest energy 
cosmic ray flux above $\sim 10^{11}$ GeV \cite{HECR}, as discussed in 
sec.\ref{sec-LR}.

\section{Conclusions}
\label{sec-concl}

In this paper, we have investigated the possibility of a solution to the GUT 
monopole problem via the so-called thermal inflation \cite{LythStew}. We first 
pointed out that the monopole problem is a problem of all unified theories of 
the strong weak and electromagnetic interaction, even of those based on non 
semi-simple gauge groups, as long as the $U(1)_Y$ gauge symmetry of the standard 
model gauge group is embedded in a non abelian gauge symmetry. Also, if there is 
a theory a 
quantum gravity which describes the universe above the Planck scale, this does 
not solve the monopole problem in the sense that monopoles are still formed 
if a unified theory exists at the grand unified scale. But this ultimate theory 
could manifest 
itselfs at low energies by introducing non renormalisable couplings, couplings 
which are suppressed by some high energy threshold. 

We have shown that thermal inflation \cite{LythStew} arises naturally in GUTs 
based on rank 
greater than five gauge groups $G$ when non-renormalisable couplings are used to 
force the GUT Higgs fields which are used to lower the rank of the group by one 
unit, to acquire a vev. We have also shown that this period of thermal inflation 
can provide enough e-foldings to solve the monopole problem. The monopoles must 
be formed before the onset of thermal inflation. Therefore 
the simplest form of the spontaneous symmetry breaking patterns which solves 
the monopole problem are given by Eq.(\ref{eq:SSB}). The period of thermal 
inflation is driven by two inflaton fields, $\Phi_{B-L}$ and 
$\overline{\Phi}_{B-L}$, which also break $U(1)_{B-L}$ when acquiring vevs at 
the end of inflation. The monopole problem is solved provided that the scale 
$M_{B-L} \geq 10^{12}$ GeV.
At the end of inflation fat $B-L$ cosmic strings are formed. These strings could 
explain the baryon asymmetry which is observed in our universe 
\cite{paperlept}. They could also explain the diffuse gamma ray 
background observed above the TeV scale, and the HECR observed above $10^{10}$ 
GeV  \cite{HECR}, if $M_{B-L} \sim 10^{14}$ GeV. If SuperKamiokande data 
correspond to $\nu_\mu \rightarrow 
\nu_{\tau}$-oscillations, if left-handed neutrinos acquire their masses via the 
see-saw mechanism \cite{seesaw} and if right-handed neutrino acquire a 
superheavy Majorana mass via the renormalisable coupling given in 
eq.(\ref{eq:NR}),  $M_{B-L} \sim 10^{14}$ GeV is also the energy scale needed to 
fit the data.

\section*{Acknowledgement}

The author would like to thank T. Barreiro for discussions. This work was 
supported by PPARC grant ${\rm n}^o$ GR/K55967.

\end{document}